\newlength{\extralineskip}
\documentclass[12pt]{article}
\usepackage{epsfig}

\begin{document}

\begin{titlepage}
\begin{flushright}
\begin{minipage}[t]{12em}
\large UAB--FT--468\\
IFT--P.053/99\\
June 1999
\end{minipage}
\end{flushright}

\vspace{\fill}

\begin{center}
\baselineskip=2.5em
{\large \bf Long
range neutrino forces in the
cosmic relic neutrino background}
\end{center}
\vspace{\fill}

\begin{center}
{\bf F. Ferrer${}^a$, J. A.
Grifols${}^a$, and M.
Nowakowski${}^b$}\\
\vspace{0.6cm}
${}^a${\em Grup de F\'\i sica
Te\`orica and Institut de F\'\i sica d'Altes
Energies\\
Universitat Aut\`onoma de Barcelona\\
08193 Bellaterra, Barcelona, Spain}\\
\vspace{0.4cm}
    ${}^b${\em Instituto de F\'{\i}sica
Te\'{o}rica\\ Universidade Estadual
  Paulista\\Rua Pamplona 145,
01405-900 S\~ao Paulo, Brazil}
\end{center}
\vspace{\fill}

\begin{center}
\large Abstract
\end{center}
\begin{center}
\begin{minipage}[t]{36em}
Neutrinos mediate long range forces among macroscopic bodies in vacuum. When the bodies are placed in the neutrino cosmic background, these forces are modified. Indeed, at distances long
compared to the scale $T^{-1}$, the relic neutrinos
completely screen
off the 2-neutrino
exchange force, whereas for small
distances the interaction remains
unaffected. 
\end{minipage}
\end{center}
\vspace{\fill}

\end{titlepage}
\clearpage

\addtolength{\baselineskip}{\extralineskip}

Dispersion potentials arising from double particle exchange have been
systematically studied
in a wide variety of physical contexts and with quite different scopes
and purposes~\cite{fs.review}. Indeed,
the studies include pure QED
phenomena such as Van der Waals
interactions~\cite{waals}, two neutrino
forces among macroscopic bodies~\cite{neutrino}, forces mediated by
scalar particles~\cite{scalar,beyond} found in recent
completions of the
standard model (e.g. superlight scalar partners of
the gravitino), etc.
In particular 2-neutrino exchange forces have been repeatedly
scrutinized
since first discussed by
Feinberg and Sucher. An aspect that
has been reanalysed in recent work~\cite{plb}
is the observation
raised
in~\cite{hp} that the cosmic neutrino heat bath has an
effect on long range
neutrino interactions. In both these papers~\cite{hp,plb} an
approximate neutrino
distribution function
was used that simplified the
calculations. The claim was that for small
neutrino chemical
potential,
the background neutrinos can be considered nearly Boltzmann
distributed
and this
fact, while only introduces a small distortion into the long
range
forces, 
makes the
calculations much easier. But the actual phase-space distribution for
relic
cosmological neutrinos has a Fermi-Dirac shape. Indeed, any
fermionic species in thermal equilibrium which at time $t_D$ and
temperature
$T_D$ of decoupling was highly relativistic followed an equilibrium
distribution $n({\mathbf{p}},t_D)=\left[ \exp \left(E/T_D \right) + 1
\right]^{-1}$. After decoupling, the energy is red shifted by the
expansion
of the Universe, $E(t)=E(t_D)\left( R(t_D)/R(t)\right)$, as the number
density decreases like $R^{-3}$. As a result, the phase-space
distribution at
time $t$ will keep the Fermi-Dirac form with the temperature $T(t)=T_D
\left(
R(t_D)/R(t)\right)$. In the present paper we use the exact Fermi-Dirac
neutrino distribution function with arbitrary chemical potential and
observe that the long distance results are drastically modified even for
small chemical potential in contrast to previous claims.  We neglect the
effect of a neutrino mass which for the
phenomenologically suggested values would not affect the present
results. We comment in
passing that there has been in the recent literature~\cite{cosmic}
renewed interest in
cosmic neutrino degeneracy which could make neutrino scattering a viable
explanation for the
Ultra-High-Energy Cosmic Ray events observed so far.

We shall adopt
the notation
in~\cite{hp,plb} and write,
\begin{equation}\label{potential}
V({\mathbf{r}})=-\int {d^{3}{\bf Q}
\over(2\pi)^{3}} \exp (i{\bf Q}\cdot {\bf r}) {\cal T}({\bf Q})
\end{equation}
where ${\cal T}({\bf Q})$ is the nucleon-nucleon elastic
scattering amplitude
(Fig. 1) in the static
limit, i.e. momentum
transfer $Q\simeq(0,{\bf Q})$, where matter is
supposed to be at
rest in
the microwave background radiation (MWBR) frame. It can be cast
in the
form 
\begin{equation}\label{amplitude} 
{\cal T}(Q)=-2iG_{F}^2
(g_{V},-2g_{A}{\bf S})^{\mu}(g'_{V},-2g'_{A}
{\bf S}')^{\nu}I_{\mu \nu}
\end{equation}
\begin{figure}[bht]
\begin{center}
\epsfig{file=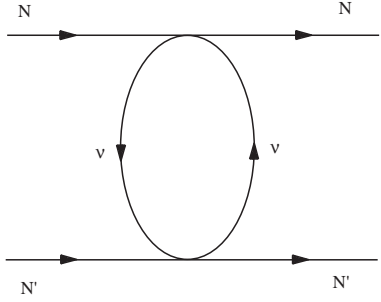,width=5cm,height=4cm}
\end{center}
\caption{{\it Lowest order Feynman diagram for two neutrino
exchange
in the four fermion effective theory.}}
\end{figure}
with
\begin{equation}\label{tensor}
 I_{\mu \nu}=\int {d^{4}k \over(2\pi)^4}
Tr[\gamma_{\mu}OiS_{T}(k)\gamma_{\nu}OiS_{T}(k-Q)]
\end{equation}
and ${\bf S},{\bf S}'$ are spin operators. The operator $O$ is the
left-handed projector ${1\over
2}(1-\gamma_{5})$ for Dirac neutrinos.
The temperature dependent
propagator $S_{T}$ has the explicit form
\begin{equation}\label{propagator}
S_{T}(k)=\rlap / k
\left[(k^2+i\epsilon)^{-1}+2\pi i\delta(k^2)(\theta
(k^0)n_{+}+\theta(-k^0)n_{-})\right]
\end{equation}
where $n_{+}$ and
$n_{-}$ are Fermi-Dirac distribution functions for
particle and
antiparticle, respectively. As discussed in~\cite{hp},
Fig. 1 evaluated
with this propagator taken together with the usual Feynman
rules is
sufficient to calculate the potential. In equation (2), $g_{V,A}$ are
composition-dependent weak vector and axial-vector
couplings. We focus
on the
spin-indepen\-dent potential, that is the $g_{V}g_{V}'$ component
of
Eq.(2). Physically this component in the potential arises because the
helicity flip produced by single neutrino
exchange is balanced  by
the
exchange of the second neutrino and, as a consequence, a
spin-independent interaction takes place that leads to a coherent effect
over many particles in
bulk matter.
Use of the first piece in equation
(3) gives the well known vacuum
result
\begin{equation}\label{feinbergsucher}
V_{0}(r)={G_{F}^2g_{V}g_{V}'\over
4\pi^3r^5}.
\end {equation}

In a neutrino background, a contribution to
the long range force can
arise
because a neutrino in the thermal bath
may be excited and de-excited
back to its
original state in the course
of the double scattering process. This
effect is
described by the
crossed terms contained in $I_{\mu\nu}$ that involve
the
thermal piece
of one neutrino propagator along with the vacuum piece of
the
other
neutrino propagator. This thermal piece of the tensor
$I_{\mu\nu}$
can
be written as
\begin{eqnarray}\label{thermal tensor}
I^{\mu
\nu}_{T}&=&-\pi
i\int{d^{4}k\over(2\pi)^4}\delta(k^2)[\theta(k^0)n_{+}+\theta(-k^0)n_{-}
]
\nonumber \\
&&\times \left[{Tr\left[\gamma^{\mu}
(\rlap /k+\rlap
/Q)\gamma^{\nu} \rlap / k\right]\over
(k+Q)^2+i\epsilon}+{Tr\left[\gamma^{\mu}\rlap / k\gamma^{\nu} (\rlap
/k-\rlap /Q)\right]\over 
(k-Q)^2+i\epsilon}\right] .
\end{eqnarray}

The temperature dependent potential
\begin{equation}\label{Tpot}
V_{T}(r)={iG_{F}^2g_{V}g_{V}'\over \pi^2r}\int_{0}^\infty
dQ\,Q\,I^{00}_{T}(Q)\,\sin Qr
\end {equation}
involves the
$I^{00}_{T}(Q)$ component: 
\begin{equation}\label{thermal tensor2}
I^{00}_{T}(Q)={-i\over \pi^2}\int
dk\,k^3\,\int_{-1}^{1}dz\,{(1-z^2)\over {4k^2z^2-Q^2}}
(n_{+}+n_{-})
\end {equation}
with $n_{\pm}(k^0\equiv \omega,T)=(e^{\omega /T \mp
\mu/T}+1)^{-1}$, $\mu$
being the chemical
potential of the neutrinos.
 
We change now the order of the integrations and perform first the
integration over $Q$, followed by the
angular integration, that is the
integration over $z$. The first step
gives,
\begin{equation}\label{Tpot2}
V_{T}(r)=-{G_{F}^2g_{V}g_{V}'\over
2\pi^3r}\int_{0}^\infty
dk\,k^3\,(n_{+}+n_{-})\int_{-1}^{1}dz\,(1-z^2)\,\cos 2kzr.
\end{equation}
The result of the $z$-integration can be cast in the form
\begin{equation}\label{Tpot3}
V_{T}(r)=-{G_{F}^2g_{V}g_{V}'\over
4\pi^3r^4}\left[1-r{d\over
dr}\right]I_{T}(r;\mu)
\end {equation}
with
\begin{equation}\label{Tpot4}
I_{T}(r;\mu)\equiv \int_{0}^\infty
d\omega\,\left((e^{\omega /T - \mu/T}+1)^{-1}+(e^{\omega /T +
\mu/T}+1)^{-1}\right)\,\sin 2\omega r.
\end {equation} 
The thermal
integral $I_{T}(r;\mu)$ can be done by realising that the
Fermi-Dirac
distribution function
can be written as an infinite series
\begin{equation}\label{fermi}
{1\over e^{x}+1}=\sum_{n=1}^{\infty}(-1)^{n+1}e^{-n \,x}
\end {equation}
Our potential is now
an infinite series where every single integration
can be easily
performed.
The final result is expressible in terms of the
hypergeometric function
$F(a,b;c;z)$. Indeed, we have
\begin{eqnarray}\label{Tpot5}
I_{T}(r;\mu)&=&{1\over
4r}\left[F(1,-2irT;1-2irT;-e^{-\mu/T})+F(1,-2irT;1-2irT;-e^{\mu/T})
\right. \nonumber \\
&+&F(1,2irT;1+2irT;-e^{-\mu/T})+F(1,2irT;1+2irT;-e^{\mu/T}) \nonumber \\
&-&\left. 8\pi rT\,\cos 2r\mu\,{\mathrm {csch}}\,2\pi rT \right], 
\end{eqnarray}
which is to be plugged into Eq.(9).

Let us single out a
few special
cases. Start with nondegenerate
neutrinos ($\mu=0$). In that
case
the
argument of the hypergeometric function is $-1$ and we may use
the following property: 
\begin {equation}\label{prop}
F(1,a;1+a;-1)={a\over
2}[\psi ({1\over 2}+{a\over 2})-\psi ({a\over 2})]
\end{equation}
where
$\psi(z)$ is the logarithmic derivative of the
$\Gamma(z)$
function. Two
further properties
of $\psi(z)$ are of help
here,   
\begin{eqnarray}\label{prop2}
{\psi({1\over 2}+z)-\psi({1\over
2}-z)}={\pi\,\tan \pi z} \nonumber\\
{\psi(z)-\psi(-z)}={-\pi\,\cot \pi
z}-{1\over z}.
\end {eqnarray}
After some straightforward algebra
$I_{T}(r;\mu=0)$ reads
\begin{equation}\label{Tpot6}
I_{T}(r;\mu=0)={1\over 2r}[1-2\pi rT\,{\rm csch}\,2\pi rT].
\end{equation}  
Finally, the temperature dependent potential for
nondegenerate relic
neutrinos is:           
\begin{equation}\label{Tpot7}
V_{T}(r)=-V_{0}(r)[1-\pi rT\,{\rm
csch}\,2\pi rT(1+2\pi rT\,\coth 2\pi rT)]
\end {equation}
where
$V_{0}(r)$ is the Feinberg Sucher potential (see Eq.(5)).
At short
distances, $r\ll 1.2\, mm$, i.e short compared to the distance
scale
set
by the neutrino
background temperature, the temperature dependent
piece
of the potential
is neglegible. It is   
\begin{equation}\label{Tpot8}
V_{T}(r)\approx -{14\over 45}V_{0}(r)(\pi
rT)^4.
\end {equation}
At
large distances (i.e. $rT\gg 1$), on the
contrary, the temperature dependent effect 
exactly cancels the vacuum
component,  
\begin{equation}\label{Tpot9}
V_{T}(r)\approx -V_{0}(r).
\end {equation}

The other instance that we wish to explore now is the
case of cold
degenerate neutrinos ($T\simeq 0, \mu
\not= 0$).
Here we
use the relation
\begin{equation}\label{rel}
F(1,0,1;z)=1
\end
{equation}
and obtain  
\begin{equation}\label{Tpot10}
I_{T\simeq
0}(r;\mu)\simeq {1\over r}(1-\cos 2\mu r).
\end {equation}
So
finally,
we have for the potential
\begin{equation}\label{Tpot11}
V_{T\simeq
0}(r)\simeq -2V_{0}(r)[1-\cos 2\mu r-\mu r\,\sin
2\mu r]
\end
{equation}
which agrees with the result given in~\cite{hp} for this
special
limit\footnote{The situation described in~\cite{hp}, namely a SN
interior, involves only a background of
neutrinos; as antineutrinos are
not retained, the result then has an
additional factor of 1/2.}. 
\begin{figure}[bht]
\begin{center}
\epsfig{file=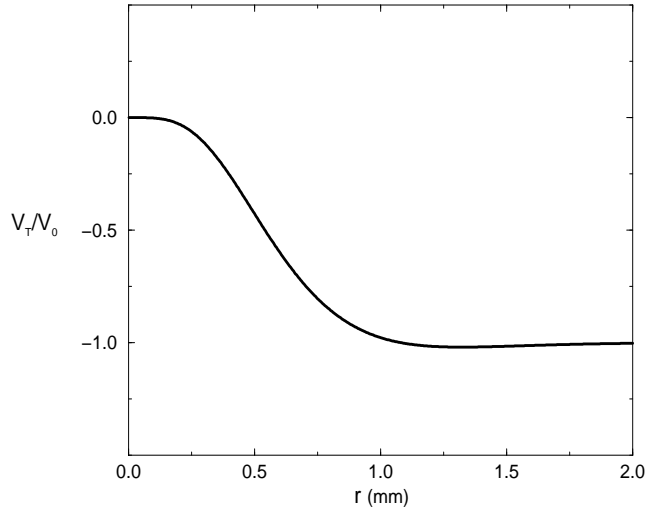,width=9cm,height=8cm}
\end{center}
\caption{\it
Ratio between the potential $V_T$, at $T \sim 1.98
  K$ corresponding to
the cosmic neutrino background, and the Feinberg-Sucher potential
$V_{0}$ when $\mu/T = 1$.}
\end{figure}
To illustrate a general
situation where we cannot make use
of
approximations, we plot the ratio
$V_{T}(r)/V_{0}(r)$ as a function
of distance for a chemical potential
on the order of the neutrino background temperature (well within the
bounds on the cosmic neutrino
degeneracy that come from primordial
nucleosynthesis~\cite{nucleo} as
well as structure formation
studies~\cite{struct}). This is depicted in
Fig.2. The curve clearly
shows the general trend: at
short distances
(much less than $1\,mm$) the
effect of the neutrino
relic background on
the neutrino
exchange
potential is smallish and for distances about
$1\,mm$ and
beyond relic
neutrinos tend to
screen it off.

We close this
paper with a brief
summary. Neutrinos mediate (very
feeble)
long-range forces between
macroscopic bodies in a vacuum. Indeed
double neutrino exchange among
matter fermions generates spin-independent forces that extend coherently
over macroscopic
distances. When the
bodies lie
in a neutrino heat bath these
forces are
altered. The
phenomenon was first studied in~\cite{hp}
and further
explored by the
present  
authors~\cite{plb}. Here we have reanalysed
the effect of a
neutrino
background on the neutrino mediated
forces
using the exact
Fermi-Dirac distribution function with arbitrary
chemical potential.
This
was not done before where, for the sake of a
simpler calculation,
either
a cold extremely degenerate (suited
e.g. for
supernova neutrinos)
neutrino sea or a Maxwell-Boltzmann
neutrino gas
were used. The present analysis has led to quite different conclusions
as to the long-range behaviour of the neutrino induced
interactions.
Indeed, the relic
neutrino background, contrary to
previous claims,
completely cancels the long distance tail ($r\geq 1\,mm$) of the
two-neutrino-exchange force and leaves  the short
distance ($r\ll
1\,mm$) component of the
interaction unaffected. Although we still lack
an experimental
confirmation of the existence of relic neutrinos in spite of many
suggestions  to detect them \cite{detection}, their theoretical status
is well established
within the Big Bang theory. Therefore their effect on neutrino mediated
long range forces is
indisputable.

\newpage

\noindent {\bf Acknowledgements} 

F.F. would like to
thank  G. Raffelt for the hospitality extended to him
during a visit to MPI, M\"unchen. F.F. also thanks  G. Raffelt and L.
Stodolsky for useful
discussions. Work partially supported by the CICYT Research Project
AEN98-1093. F.F.
acknowledges the CIRIT for financial support.
M.N. would like to thank
Funda\c{c}\~ao
de Amparo \`a Pesquisa de 
S\~ao Paulo (FAPESP) and
Programa de Apoio a
N\'ucleos de Excel\^encia
(PRONEX). 
\vskip 2cm
%

\end{document}